\documentclass[epj,referee]{svjour}
\usepackage{graphics,epsfig}
\newcommand{\M}{{\cal M}}
\newcommand{\MeV}{{\rm MeV}}
\newcommand{\GeV}{{\rm GeV}}

\newcommand{\fm}{{\rm fm}}

\renewcommand{\Im}{{\rm Im}}

\begin{document}
\title{Model independent constraints from vacuum and in-medium QCD Sum Rules
}
\author{F.~Klingl and W.~Weise 
\thanks{Work supported in part by BMBF and GSI}
}
\offprints{weise@physik.tu-muenchen.de}          
% Insert a name or remove this line
%
\institute{Physik-Department, Theoretische Physik, Technische Universit\"at
M\"unchen, D-85747 Garching, Germany}
\date{Received: date / Revised version: date}
\abstract{We discuss QCD sum rule constraints based on moments of vector meson
  spectral distributions in the vacuum and in a nuclear medium. Sum rules for
  the two lowest moments of these spectral distributions do not suffer from
  uncertainties related to QCD condensates of dimension higher than four. We
  exemplify these relations for the case of the $\omega$ meson and discuss
  the issue of in-medium mass shifts from this viewpoint.
 % end of PACS codes
} %end of abstract
\maketitle
\label{intro}
QCD sum rules have repeatedly been used in recent times to arrive at estimates
for possible in-medium mass shifts of vector mesons \cite{2,7}. The validity of
such estimates has been questioned, however, for several reasons. First,
for broad structures such as the $\rho$ meson whose large vacuum
decay width is further magnified by in-medium reactions, the QCD sum rule
analysis does not provide a reliable framework to extract anything like a ``mass
shift'' \cite{4,5}. Secondly, notorious uncertainties exist at the level of
factorization assumptions commonly used to approximate four-quark condensates
in terms of $\langle \bar{q} q \rangle^2$, the square of the standard chiral
condensate. The first objection is far less serious for the $\omega$ meson
which may well have a chance to survive as a reasonably narrow quasi-particle
in nuclear matter \cite{4,4b}. The second objection, however, is difficult to overcome:
factorization of four-quark condensates may indeed be questionable.

In the present note we focus on the two lowest moments ($\int ds s^n R(s)$ 
with $n = 0,1$) of vector
meson spectral distributions, in vacuum as well as in nuclear matter, 
and point out that these are subject to sum rules which do {\cal not} 
suffer from the uncertainties introduced by
four-quark condensates. These sum rules are shown to provide useful, model
independent constraints which we exemplify for the case of the $\omega$ meson
spectral distribution and its change in the nuclear medium. The sum rule for
the second moment, $\int ds s^2 R(s)$, does involve the four-quark condensate.
In fact it can be used in principle to determine this particular condensate
and test the factorization assumption. The detailed analysis of this 
question will be defered to a longer paper. In this short note we confine
ourselves to conclusions that can be drawn without reference to four-quark
condensates. 

The starting point is the current-current correlation function
 \begin{equation}
  \Pi_{\mu\nu}(q)=i\int d^4x \: e^{iq\cdot x}\langle {\cal T}j_{\mu}(x) j_{\nu}(0)\rangle  
 \label{2.1}
\end{equation}
where ${\cal T}$ denotes the time-ordered product and the expectation value is
taken either in the vacuum or in the ground state of nuclear matter at rest. In
vacuum the polarization tensor (\ref{2.1}) can be reduced to a single scalar
correlation function, $\Pi(q^2)={\scriptstyle \frac{1}{3}} g^{\mu \nu} \Pi_{\mu \nu} (q)$. In
nuclear matter there are two (longitudinal and transverse) correlation
functions which coincide for a meson at rest with respect to the medium
(i.e. with $q^\mu=(\omega,\vec{q}=0)$). 

The reduced correlation function is written as a (twice subtracted) dispersion relation,
\begin{equation}
  \Pi(q^2)=\Pi (0)+\Pi'(0) \, q^2+\frac{q^4}{\pi}\int ds\frac{\Im\Pi (s)}{s^2(s-q^2
  -i\epsilon)}.
\label{2.14}
\end{equation}
where  $\Pi (0)$ vanishes in vacuum but contributes in nuclear matter. At large spacelike $Q^2=-q^2>0$ the QCD
operator product (Wilson) expansion gives
\begin{equation}
  \label{2.14a}
  12 \pi^2 \Pi(q^2=-Q^2) = -c_0 Q^2
  \ln{\left(\frac{Q^2}{\mu^2}\right)}+ c_1
  +\frac{c_2}{Q^2}+\frac{c_3}{Q^4}+... 
\end{equation}
We specify the coefficients $c_i$ for the isoscalar current
$j^\mu={\scriptstyle \frac{1}{6}}(\bar{u} \gamma^\mu u +\bar{d}\gamma^\mu d)$,
the case of the $\omega$ meson that we wish to use here for explicit
evaluations. In vacuum we have:
\begin{eqnarray}
  \label{2.15}
c_0 &=& \frac{1}{6} \left(1+\frac{\alpha_S}{\pi} \right)
,\hspace*{1cm} c_1=-\frac{1}{2}(m_u^2+m_d^2), \\ 
c_2&=& \frac{\pi^2}{18} \langle \frac{\alpha_S}{\pi} {\cal G}^{\mu \nu}{\cal
G}_{\mu \nu}\rangle + \frac{2 \pi^2}{3} \langle m_u u\bar{u}+ m_d d\bar{d}
 \rangle,\label{2.15e}
\end{eqnarray}
while $c_3$ involves combinations of four-quark condensates of (mass) dimension
6. The quark mass term $c_1$ is small and can be dropped in the actual
   calculations. For the gluon condensate we use $\langle \frac{\alpha_S}{\pi}
   {\cal G}^2\rangle=(0.36\, \GeV)^4$ \cite{9}, and the (chiral) quark condensate is
   given by $\langle  m_u \bar{u}u+ m_d \bar{d}d  \rangle \simeq m_q\langle 
   \bar{u}u+ \bar{d}d  \rangle=-m_\pi^2 f_\pi^2\simeq-(0.11 \, \GeV)^4$ through
   the Gell-Mann, Oakes, Renner relation.

In the nuclear medium with baryon density $\rho$ we have
$c_i(\rho)=c_i(\rho=0)+\delta c_i(\rho)$ with $c_i(0)$ given by
eqs.(\ref{2.15},\ref{2.15e}), and 
\begin{equation}
\delta c_2(\rho)=\frac{\pi^2}{3} \left[-\frac{4}{27} M_N^{(0)}+2 \sigma_N+ A_1
\, M_N \right] \rho
\label{5.7c}
\end{equation}
to linear order in $\rho$. The first term in brackets is the leading
density dependent correction to the gluon condensate and
involves the nucleon mass in the chiral limit, $M_N^{(0)}\simeq 0.75 \, \GeV$
\cite{6}. The second part proportional to the nucleon sigma term $\sigma_N
\simeq 45 \, \MeV$ is the first order correction of the quark
condensate, and the third term introduces the first moment of the quark
distribution function in the nucleon:
\begin{equation}
  \label{hah}
  A_1= 2 \int dx \,x \, \left[ u(x)+\bar{u}(x)+d(x)+\bar{d}(x) \right] .
\end{equation}
It represents twice the fraction of momentum carried by quarks in the proton. We
take $A_1 \simeq 1$ as determined by deep-inelastic lepton scattering at $Q
\sim 2 \, \GeV$. Note that $\delta c_2(\rho_0) \simeq 4 \cdot 10^{-3} \,
\GeV^4$ at $\rho=\rho_0=0.17 \, \fm^{-3}$, the density of nuclear matter, and
almost all of this correction comes from the term proportional to $A_1$.

Next we introduce the Borel transform of eq.~(\ref{2.14a}):
\begin{equation}
  \label{2.20} 
  12 \pi^2 \Pi(0) +\int_0^{\infty}ds\, R (s)
  e^{-s/ \M^2} = c_0 \M^2 +c_1+\frac{c_2}{\M^2}+\frac{c_3}{2 \M^4}+...
\end{equation}
with $ R(s)= -\frac{12 \pi}{s} \Im \Pi(s)$ and $\Pi(0)=-\rho/4 M_N$, the
vector meson analogue of the Thomson term in photon scattering.

We separate the spectrum $R(s)$ into a resonance part with $s \leq s_0$ and a
continuum $R_c(s)$ which must approach the perturbative QCD limit for $s>s_0$:
\begin{equation}
  R_c(s) = \frac{1}{6} \left( 1+\frac{\alpha_S}{\pi} \right) \Theta (s- s_o).
\label{2.13}
\end{equation}
The factor $\scriptstyle \frac{1}{6}$ is again specific for the isoscalar
channel. The Borel mass parameter $\M$ must be sufficiently large so that
eq.(\ref{2.20}) converges rapidly, but otherwise it is arbitrary. We choose $\M
>\sqrt{s_0}$ so that $e^{-s/\M^2}$ can be expanded in powers of $s/\M^2$ for
$s<s_0$. The remaining integral $\int_{s_0}^\infty ds \, R_c(s) e^{-s/\M^2}$ is
evaluated inserting the running coupling strength $\alpha_S(s_o)$ in
eq.(\ref{2.13}). Then the term-by-term comparison in eq.(\ref{2.20}) gives
the following set of sum rules for the moments of the spectrum $R(s)$ (see also
refs. \cite{7,8})
\begin{eqnarray}
\label{c1}
 \int_0^{s_0}ds\, R (s) & = &s_0 c_0+c_1 -12 \pi^2 \Pi(0), \\
   \int_0^{s_0}ds\, s \,  R (s) & =& \frac{s_0^2}{2}  c_0 -c_2, \label{c2} \\
 \int_0^{s_0}ds\, s^2 \,  R (s) & = &\frac{s_0^3}{3}  c_0 +c_3.\label{c3}
\end{eqnarray}
Note that the first two sum rules are well determined and represent useful
constraints for the spectrum $R(s)$. Only the third sum rule (\ref{c3})
involves four-quark condensates which are uncertain. In this short paper
we concentrate on eqs. (\ref{c1},\ref{c2}). A detailed analysis of 
eq.(\ref{c3}) will be presented in a forthcoming longer paper.
It is instructive to illustrate the sum rules (\ref{c1},\ref{c2}) for the
$\omega$ meson in vacuum using Vector Meson Dominance (VMD) for the resonant
part of $R(s)$. In this model we have
\begin{equation}
  R(s) = 12 \pi^2 \frac{m_\omega^2}{g_\omega^2} \, \delta(s-m_\omega^2)+ \frac{1}{6} \left( 1+\frac{\alpha_S}{\pi} \right) \Theta (s- s_o).
\label{vmd}
\end{equation}
with $g_\omega=3g\simeq 16.8$ (using the vector coupling constant $g=5.6$). We can
neglect the small quark mass term $c_1$ and find from eq.~(\ref{c1}):
\begin{equation}
  \label{vmd2}
  \frac{8 \pi^2}{g^2} \frac{m_\omega^2}{s_0}= 1+\frac{\alpha_S}{\pi},
\end{equation}
which fixes $\sqrt{s_0}=1.16 \, \GeV$ using $\alpha_S(s_0) \simeq 0.4$ and
$m_\omega=0.78 \, \GeV$. It is interesting to identify the spectral gap
$\Delta = \sqrt{s_0}$ with the scale for spontaneous chiral symmetry 
breaking, $\Delta = 4\pi f_{\pi}$, where $f_{\pi} = 92.4~MeV$ is the pion
decay constant. In the VMD model, taking the zero width limit,
eq.(14) holds for both the $\omega$ and $\rho$ meson, with equal
mass $m_V = m_{\rho} = m_{\omega}$. Inserting $s_0 = 16\pi^2 f_{\pi}^2$ in
eq.(14) one recovers the famous KSFR relation $m_V = \sqrt{2}~gf_{\pi}$ up
to a small QCD correction.

The sum rule (\ref{c2}) for the first moment gives
\begin{equation}
  \label{vmd3}
  \frac{8 \pi^2}{g^2} m_\omega^4= \frac{s_0^2}{2}
  \left(1+\frac{\alpha_S}{\pi} \right) - \frac{\pi^2}{3} \left[\langle
  \frac{\alpha_S}{\pi} {\cal G}^2\rangle + 12 \langle  m_u \bar{u}u+ m_d
  \bar{d} d \rangle\right].
\end{equation}
Inserting the values for the gluon and quark condensates we find indeed
perfect consistency. 
Given a model for the $\omega$ meson spectral function in the vacuum and in the
nuclear medium, the sum rules (\ref{c1},\ref{c2}) therefore provide useful
constraints to test the calculated spectra.

\begin{figure*}
\unitlength1mm
\begin{picture}(150,75)
\put(0,0){\makebox{\epsfig{file=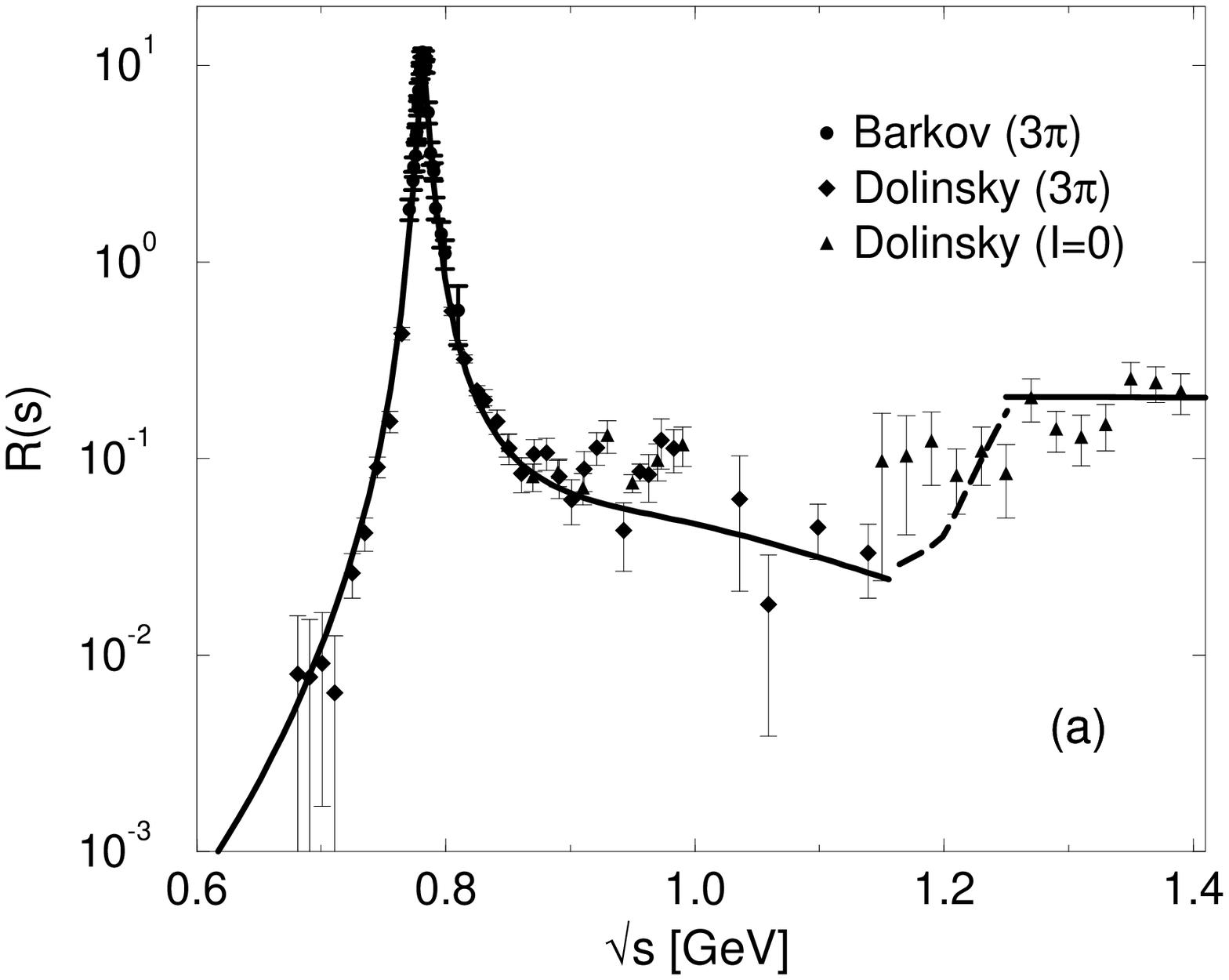,width=84mm}}}
\put(80,0){\makebox{\epsfig{file=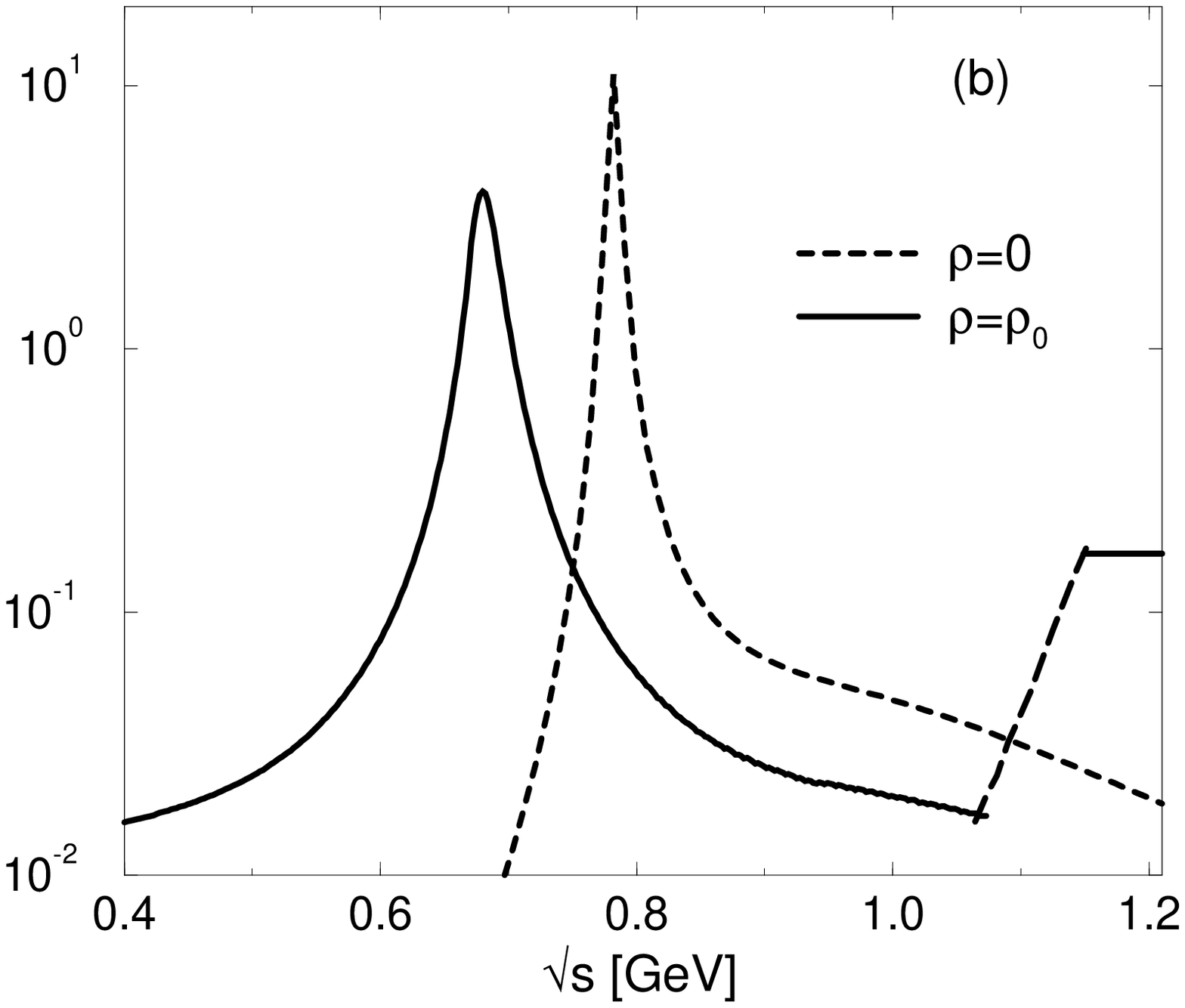,width=84mm}}}
\end{picture}
\vspace*{-4mm}
\caption{a) Spectrum $R(s)$ in the $\omega$ meson channel as calculated in
ref.~\protect \cite{4} (solid line). The data points refer to $e^+e^- \to 3
\pi$ and $e^+e^- \to$ hadrons $(I=0)$ \protect \cite{10}
\protect \newline b) In-medium spectrum of $\omega$ meson excitations in
nuclear matter at density $\rho_0=0.17 \, \fm^{-3}$ as calculated in
refs.\protect \cite{4,4b} (solid line) in comparison with the vacuum spectrum
(dashed line).}
\label{fig1}       % Give a unique label
\end{figure*}

We now continue on from VMD to a more realistic approach. In refs.~\cite{4,4b}
we have used an effective Lagrangian based on chiral $SU(3)\otimes SU(3)$
symmetry with inclusion of vector mesons as well as anomalous couplings from
the Wess-Zumino action in order to calculate the $\omega$ meson spectrum both
in the vacuum and in nuclear matter. The resulting vacuum spectrum reproduces
the observed $e^+e^- \to~$hadrons$(I=0)$ data very well \cite{4} (see
Fig.~\ref{fig1}a). The predicted in-medium mass spectrum (for $\omega$ excitations with
$\vec{q}=0$) shows a pronounced downward shift of the $\omega$-meson peak and a
substantial, but not overwhelming increase of its width from reactions 
such as $\omega N
\to \pi N,\, \pi\pi N$ etc. (see Fig.~\ref{fig1}b). At large $s>s_0$, both spectra should
approach the QCD limit (\ref{2.13}).
The consistency test of these calculated spectral distributions with the sum
rules (\ref{c1}) and (\ref{c2}) goes as follows:
\begin{itemize}
\item \underline{the vacuum case}: \newline
the two sides of eq.~(\ref{c1}),
\begin{equation}
 \label{vaccase}
\int_0^{s_0}ds\, R(s) = \frac{s_0}{6} \left( 1+\frac{\alpha_S(s_0)}{\pi} \right),
\end{equation}
now match at $\sqrt{s_0}=1.25 \, \GeV$, with $ \int_0^{s_0}ds\, R(s)=0.29 \,
\GeV^2$. The sum rule for the first moment gives  $ \int_0^{s_0}ds\,s\,
R(s)=0.19 \,\GeV^4$, to be compared with  $ \frac{1}{2}
s_0^2(1+\alpha_s/\pi)-c_2=0.22 \, \GeV^4$, so there is consistency at the 10\%
level.
\item \underline{the in-medium case}: \newline
now we have to match the moments of the density dependent spectral
distributions,
\begin{equation}
 \label{medcase}
\int_0^{s_0}ds\, R(s,\rho) = \frac{s_0}{6} \left( 1+\frac{\alpha_S(s_0)}{\pi}
\right)+\frac{3 \pi^2}{M_N} \rho,
\end{equation}
together with
\begin{equation}
 \label{medcase2}
\int_0^{s_0}ds\,s\, R(s,\rho) = \frac{s_0^2}{12} \left( 1+\frac{\alpha_S(s_0)}{\pi}
\right)-c_2(0)-\delta c_2(\rho).
\end{equation}
Using our calculated spectrum \cite{4b} shown in Fig.~\ref{fig1}b, we find $\sqrt{s_0}=
1.08\,\GeV$ at $\rho=\rho_0=0.17\, \fm^{-3}$, with $\int_0^{s_0}ds\,
R(s,\rho_0) =0.26 \, \GeV^2$. Then  $\int_0^{s_0}ds\,s\,
R(s,\rho_0) =0.11 \, \GeV^4$ is to be compared with the right hand side of
eq.~(\ref{medcase2}) which gives $0.12\, \GeV^4$, so there is again excellent
consistency.

Note again that these tests do not involve uncertain four-quark
condensates. Furthermore, if the in-medium spectrum shows a reasonably narrow
quasi particle excitation, the quantity $\bar{m}^2=\int_0^{s_0}ds\,s\,
R(s)/\int_0^{s_0}ds\, R(s)$ can indeed be interpreted as the square of an
in-medium ``mass'' of this excitation. For our $\omega$ meson case we find
$\bar{m}=0.65 \, \GeV$ at $\rho=\rho_0$, a substantial downward mass shift as
discussed in refs.~\cite{4,4b}. (For the broad $\rho$ meson spectrum, on the
other hand , the interpretation of $\bar{m}$ as an in-medium mass is not
meaningful as demonstrated in ref.~\cite{4}).

Amusingly, the spectral gap $\Delta = \sqrt{s_0}$ decreases by about 15 percent
when replacing the vacuum by nuclear matter. This is in line with the 
proposition that this gap reflects the order parameter for
spontaneous chiral symmetry breaking and scales like the pion decay constant 
$f_{\pi}$ (or, equivalently, like the square root of the chiral condensate
$\langle \bar{q} q \rangle$).  

In summary, we have shown that the combination of sum rules (\ref{c1})
and (\ref{c2}) for the lowest moments of the spectral distributions does serve
as a model-independent consistency test for calculated spectral functions.
\end{itemize}

\end{document}